\documentclass[preprint,12pt]{elsarticle}
\usepackage[margin=2.4cm]{geometry}
\usepackage{graphicx}
\usepackage{multirow}
\usepackage{amsmath}
\usepackage{listings}
\usepackage{tikz}\usetikzlibrary{decorations.markings}

\newtheorem{thm}{Theorem}

\newtheorem{lem}{Lemma}

\begin{document}

\begin{frontmatter}

\title{A Similarity Solution of Rear Stagnation-point Flow over a Flat Plate in Two Dimensions}
\author{Chio Chon Kit}
\date{\today}

\begin{abstract}
This paper investigates the nature of the development of vortex shedding of two-dimensional unsteady flow of an incompressible fluid at the rear stagnation point. 
\end{abstract}


\end{frontmatter}

\section{Introduction}
The classical two-dimensional steady stagnation-point flow on th plane boundary $y=0$ can be analysed exactly by Hiemenz \cite{hiemenz1911grenzschicht}. At a rear stagnation point, on a circular cylinder say, the external flow is extracted away from the rear stagnation point. Common observation shows that when the flow is everywhere irrotational. A vortex sheet is formed near the plane and reversed flow develops in the region of vortical flow. 

Forward stagnation point at which a balance is achieved between diffusion of vorticity and the inertia results in a steady solution.  Rear stagnation-point flows against an infinite flat wall do not have analytic solution in two dimensions, but certain reverse flows have solution in three dimensions \cite{davey1961boundary}.

Proudman and Johnson \cite{proudman1962boundary} first suggested that the convection terms dominate in considering the inviscid equation in the body of the fluid. By introducing a very simple function of a particular similarity variable and neglecting the viscous forces in their analytic result for region sufficiently far from the wall, they obtained an asymptotic solution in reversed stagnation-point flow, describing the development of the region of separated flow for large time $t$.

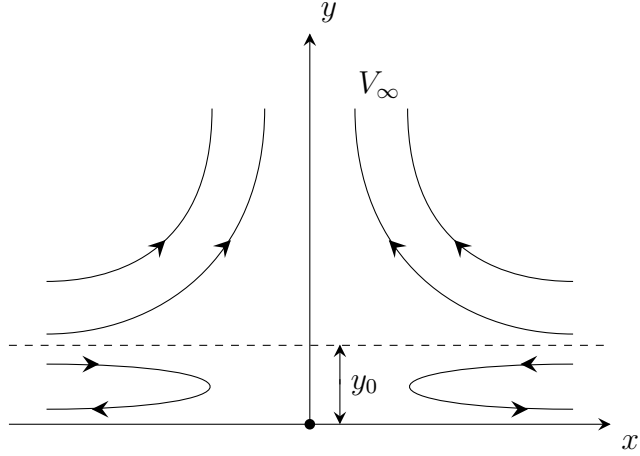
\begin{figure}[!htb]\centering
\begin{tikzpicture}[>=stealth]
\draw[
    decoration={markings,mark=at position 1 with {\arrow[scale=1.5]{>}}},
    postaction={decorate},
    shorten >=0.4pt
    ] (-4,-1.2) -- (4,-1.2);
\coordinate [label=-45:$x$] (a) at (4,-1.2);
\draw[
    decoration={markings,mark=at position 1 with {\arrow[scale=1.5]{>}}},
    postaction={decorate},
    shorten >=0.4pt
    ](0,-1.2) -- (0,4);
\coordinate [label=45:$y$] (a) at (0,4);
\coordinate [label=45:$V_\infty$] (a) at (0.5,3);
\fill[black] (0,-1.2) circle (2pt);

\draw (0.6,3) .. controls (0.6,1) and (2,0) .. (3.5,0);
\draw[
    decoration={markings,mark=at position 1 with {\arrow[scale=2]{>}}},
    postaction={decorate},
    shorten >=0.4pt
    ] (1.15,1.15) -- (1.05,1.25);
\draw (-0.6,3) .. controls (-0.6,1) and (-2,0) .. (-3.5,0);
\draw[
    decoration={markings,mark=at position 1 with {\arrow[scale=2]{>}}},
    postaction={decorate},
    shorten >=0.4pt
    ] (-1.15,1.15) -- (-1.05,1.25);
\draw (1.3,3) .. controls (1.3,1) and (2.7,0.7) .. (3.5,0.7);
\draw[
    decoration={markings,mark=at position 1 with {\arrow[scale=2]{>}}},
    postaction={decorate},
    shorten >=0.4pt
    ] (2.02,1.15) -- (1.92,1.25);
\draw (-1.3,3) .. controls (-1.3,1) and (-2.7,0.7) .. (-3.5,0.7);
\draw[
    decoration={markings,mark=at position 1 with {\arrow[scale=2]{>}}},
    postaction={decorate},
    shorten >=0.4pt
    ] (-2.02,1.15) -- (-1.92,1.25);

\draw (3.5,-0.4) .. controls (0.6,-0.4) and (0.6,-1) .. (3.5,-1);
\draw[
    decoration={markings,mark=at position 1 with {\arrow[scale=2]{>}}},
    postaction={decorate},
    shorten >=0.4pt
    ] (2.9,-0.4) -- (2.8,-0.4);
\draw[
    decoration={markings,mark=at position 1 with {\arrow[scale=2]{>}}},
    postaction={decorate},
    shorten >=0.4pt
    ] (2.8,-1) -- (2.9,-1);
\draw (-3.5,-0.4) .. controls (-0.6,-0.4) and (-0.6,-1) .. (-3.5,-1);
\draw[
    decoration={markings,mark=at position 1 with {\arrow[scale=2]{>}}},
    postaction={decorate},
    shorten >=0.4pt
    ] (-2.9,-0.4) -- (-2.8,-0.4);
\draw[
    decoration={markings,mark=at position 1 with {\arrow[scale=2]{>}}},
    postaction={decorate},
    shorten >=0.4pt
    ] (-2.8,-1) -- (-2.9,-1);
\draw [dashed] (-4,-0.15) -- (4,-0.15);

\draw[
    decoration={markings,mark=at position 1 with {\arrow[scale=1.5]{>}}},
    postaction={decorate},
    shorten >=0.4pt
    ] (0.4,-0.67) -- (0.4,-0.14);

\draw[
    decoration={markings,mark=at position 1 with {\arrow[scale=1.5]{>}}},
    postaction={decorate},
    shorten >=0.4pt
    ]  (0.4,-0.6)--(0.4,-1.2) ;
\coordinate [label=-45:$y_0$] (a) at (0.4,-0.4);
\end{tikzpicture}
\caption{Streamlines of rear stagnation-point flow}
\label{vort}
\end{figure}

The general feature of the predicted streamline pattern is sketched in Figure \ref{vort}.

\section{Flow Analysis Model}
We shall demonstrate the flow analysis for the two-dimensional case. We begin with writing the governing equations in conservative velocity form in the Cartesian coordinates:
\begin{subequations}
 \begin{gather}
\frac{\partial u}{\partial x}+\frac{\partial v}{\partial y}=0
\label{eq:e0}
  \end{gather}
 \begin{gather}
    \frac{\partial u}{\partial t}+ u \frac{\partial u}{\partial x}+v\frac{\partial u}{\partial y}=  -\frac{1}{
     \rho}\frac{\partial p}{\partial x}+ \nu \left(\frac{\partial^2 u}{\partial
     x^2}+\frac{\partial ^2u}{\partial y^2}\right)
     \label{e3}\\\notag\\
     \frac{\partial v}{\partial t}+ u \frac{\partial v}{\partial x}+v\frac{\partial v}{\partial y}=  -\frac{1}{
     \rho}\frac{\partial p}{\partial y}+ \nu \left(\frac{\partial^2 v}{\partial
     x^2}+\frac{\partial ^2v}{\partial y^2}\right)
     \label{e4}
  \end{gather}
  \label{e3_4}
\end{subequations}
The equation of continuity (\ref{eq:e0}) is integrated by introducing the stream function $\psi$:
  \begin{equation}
   u=   \displaystyle \frac{\partial \psi}{\partial y}\qquad\mathrm{and}\qquad v= \displaystyle- \frac{\partial \psi}{\partial x}
   \label {stream}
  \end {equation}
In rear stagnation flow without friction (ideal fluid flow), the stream function may be written as
  \begin{equation}
\psi =\psi_\infty = -A_\infty xy
   \label {eq:e00}
\end {equation}
where $A_\infty$ is a constant and from which
  \begin{equation}
U_\infty=-A_\infty x \qquad\mathrm{and}\qquad V_\infty= A_\infty y.
   \label {eq:e01}
 \end {equation}
We have $U_\infty=0$ at $x=0$ and $V_\infty=0$ at $y=0$, but the no-slip boundary at wall $(y=0)$ cannot be satisfied.

In a (real) viscous fluid the flow motion, Proudman and Johnson \cite{proudman1962boundary} model the flows by considering a very simple function of a particular similarity variable
\begin{subequations}
\begin{gather}
\psi = -\sqrt{A\nu}xf(\eta, \tau)\label{psi1}\\
\eta = \sqrt{\frac{A}{\nu}}y\\
\tau = At
\end{gather}
 \label{e9}
\end{subequations}
Equation (\ref{e3_4}) then gives, for $f(\eta,\tau)$,
   \begin{equation}
    f_{\eta\tau}-(f_{\eta})^2+ff_{\eta\eta}-f_{\eta\eta\eta}=-1,
     \label{e10}
  \end{equation}
with the boundary conditions
  \begin{subequations}
     \begin{gather}
        f(0,\tau)= f_{\eta}(0,\tau)=0\\
        f_{\eta}(\infty,\tau) = 1.~~~~~~~~~~
    \end{gather}
 \label{e11}
  \end{subequations}

Considering unsteady similarity variables (\ref{e9}) in the form of $\eta$ only, we have
\begin{subequations}
\begin{gather}
\psi = -\sqrt{A(t)\nu}xf(\eta, \tau)\label{psi1}\\
\eta = \sqrt{\frac{A(t)}{\nu}}y
\end{gather}
 \label{e12}
\end{subequations}
Recall the governing equation (\ref{e3})
$$
  \frac{\partial u}{\partial t}+ u \frac{\partial u}{\partial x}+v\frac{\partial u}{\partial y}=  -\frac{1}{
     \rho}\frac{\partial p}{\partial x}+ \nu \left(\frac{\partial^2 u}{\partial
     x^2}+\frac{\partial ^2u}{\partial y^2}\right)
$$
and unsteady similarity variables (\ref{e9}) yields
$$
  -\frac{1}{2} \dot{A}x\eta f'' - \dot{A}xf'+ A^2x(f')^2 - A^2xff'' 
=  -\frac{1}{\rho}\frac{\partial p}{\partial x}- A^2xf'''
$$
or 
\begin{equation}
-f'''+ff''-(f')^2 = -\frac{\dot{A}}{A^2} \left(f' +\frac{1}{2}\eta f''\right) + \frac{1}{A^2x\rho}\frac{\partial p}{\partial x}
\end{equation}
Substitution in boundary conditions (\ref{e11}) gives the equation
  \begin{subequations}
     \begin{gather}
f'''-ff''-1+(f')^2 = \kappa \left(f' +\frac{1}{2}\eta f''-1\right) \\
        f(0)= f'(0)=0\\
        f'(\infty) = 1~~~~~~~~
    \end{gather}
 \label{e13}
  \end{subequations}
where 
\begin{equation}
\kappa = \frac{\dot{A}}{A^2} 
 \label{e14}
\end{equation}
is required to be constant in time, and then integrates to give 
 \begin{equation}
A(t) = \frac{1}{\kappa(t_0 -t)} 
 \label{e15}
\end{equation}
approaching a singularity at finite time $t = t_0$.

Noting that $\kappa ={\dot{A}}/{A^2} = fL/U_{\infty}$, which is equivalent to Strouhal number. The Strouhal number describes the ratio between inertial forces due to the local acceleration and those due to the convective acceleration in unsteady flow.  Here, $f$ is the frequency of vortex shedding, $L$ is the characteristic length and $U_{\infty}$ is the external flow velocity. 

It should be noted that the dimensionless velocity distribution $f_{\eta}$ is, from (\ref{e9}), independent of the length $x$, and thus equation  (\ref{e13}) is a similarity equation of the full Navier-Stokes equation at two-dimension rear stagnation point. 

\section{Insolubility when $\kappa=0$}
It is proven that all of the solutions, however,  do not satisfy the boundary conditions when $\kappa=0$.

\begin{lem}
No solution  $f'(\eta)$ exists which has stationary value of 1 for finite $\eta$ when $\kappa=0$.
 \label{l1}
\end{lem}
Proof. Set $\kappa=0$ and rearrange Eq.~(\ref{e13}) yields\\
 \begin{equation}     
  f'''=1-(f')^2+ff''    
 \label{leq:e13}
   \end{equation}
Suppose for $\eta = \eta_0$, we have 
$f'(\eta_0)=1$ and $f''(\eta_0)=0$. Then, it follows from the derivatives of Eq.~(\ref{leq:e13}) that $f'''$ and all higher derivatives are zero when $\eta = \eta_0$.  Considering a variable transformation
\begin{gather}
\lambda(\eta)=f'(\eta)\notag\\
\lambda(\eta_0)=1
\end{gather}
Expanding the function into Taylor's series near $\eta_0$, we have
\begin{eqnarray*}
f'(\eta) = \lambda(\eta)
& = &\sum_{n=0}^{\infty} \frac{\lambda^{(n)}(\eta_0)}{n!}(\eta - \eta_0)^n \\
& = &\lambda(\eta_0)+\sum_{n=1}^{\infty} \frac{\lambda^{(n)}(\eta_0)}{n!}(\eta - \eta_0)^n \\
& = & 1+\sum_{n=1}^{\infty} \frac{\lambda^{(n)}(\eta_0)}{n!}(\eta - \eta_0)^n \\
& \equiv & 1
\end{eqnarray*}
Hence, the boundary condition $f'(0)=0$ is thus not satisfied and the Lemma is proved.

\begin{lem}
When $f'$ has a stationary value, if $|f'|<1$ it is a minimum and if $|f'|>1$ it is a maximum.
 \label{l2}
\end{lem}
Proof: From Eq.~(\ref{leq:e13}), when $f'$ has a stationary value, it means $f''=0$ and Eq.~(\ref{leq:e13}) becomes
 \begin{equation}     
  f'''=1-(f')^2    
 \label{leq:e14}
   \end{equation}
If $|f'|<1$, $f'''>0$ and it is minima. Else if $|f'|>1$, $f'''<0$ and it is maxima. Eventually, the lemma is proved.

\begin{thm}
Given any $f'(\eta) \to 1$ as $\eta \to \infty$, no solution of Eq.~(\ref{e13}) exists when $\kappa=0$.
 \label{t1}
\end{thm}
Proof : When $|f'|<1$, since $f' \to 1$ as $\eta \to \infty$, then $f''$ must be greater than zero. Hence, recall from Eq.~(\ref{leq:e13}),
$$  f'''=1-(f')^2+ ff'' >0.$$ 
for all $\eta>\eta_0$. After integrating $f'''(\eta) > 0$ from $\eta_0$ to $\eta > \eta_0$, we have 
$$f'' (\eta) > f'' (\eta_0 ) = K > 0.$$
Another integration from $\eta_0$ to $\eta > \eta_0$ yields 
$$f' (\eta) > f' (\eta_0 )+ K (\eta - \eta_0 ).$$
By Lemma $2$, $f'(\eta)$ has at most one stationary value because one cannot have two consecutive stationary values which are both minima. Since $f''(\eta)>0$, when $\eta \to \infty$, $f'(\eta) \to \infty.$ It violates that $f'(\eta) \to 1$. A similar argument shows that a solution cannot approach to 1 when $|f'|>1$. 

\section{Similarity Analysis}
Equation (\ref{e13}) is a third-order nonlinear ordinary differential equation. A crucial step in obtaining an analytical solution involves rearranging the equation as an autonomous differential equation. In mathematics, an autonomous differential equation is a system of ordinary differential equations which does not explicitly depend on the independent variable.

In order to omit the variable $\eta$ in the differential equation (\ref{e13}), it is generally accepted to make a change of variable
  \begin{equation}
  F= f+\frac{\kappa}{2}\eta
  \end{equation}
and the equation becomes 
  \begin{subequations}
     \begin{gather}
F'''-FF''+(F')^2-2\kappa F' +\frac{3}{4}\kappa^2-1+\kappa =0 \\
        F(0)= 0, ~F'(0)=\frac{\kappa}{2}.
    \end{gather}
 \label{e3_0}
  \end{subequations}
Noting that the equation (\ref{e3_0}) becomes to an autonomous differential equation when $\frac{3}{4}\kappa^2-1+\kappa =0$. In another word, if $\kappa=-2$ or $\kappa=\frac{2}{3}$, the equation (\ref{e3_0}) could be expressed into a differential equation without dependent variable. 

\subsection{Case of $\kappa=\frac{2}{3}$}
The differential equation (\ref{e3_0}) changes into the form as
  \begin{equation}
F'''-FF''+(F')^2-\frac{4}{3}F'=0
 \label{e3_7}
  \end{equation}
The solution of differential equation (\ref{e3_7}) can be obtained similarly by rewriting as another polynomial 
  \begin{subequations}
     \begin{gather}
F=a+be^{-\delta \eta} \\
       a=\frac{1}{\delta}\\
       b=-\frac{1}{\delta}
    \end{gather}
 \label{e3_8}
  \end{subequations}
Substituting (\ref{e3_8}) into equation (\ref{e3_7}), and the comparison of the coefficients in the powers of $e^{-\delta \eta}$ leads to the determination of $\delta = \pm 1$. This results into another velocity function as
  \begin{subequations}
     \begin{gather}
 f(\eta)= -\frac{1}{3}\eta \pm \frac{1}{3}(1 - e^{\mp \eta}) \\
  f'(\eta)=-\frac{1}{3}(1 - e^{\mp \eta})
    \end{gather}
 \label{e3_9}
  \end{subequations}
where $f'$ tends exponentially to a negative constant as $\eta \rightarrow \infty$. We get an immediate contradiction that the flow field is not able to remain unchanged at sufficient distances far away from the wall at any finite time. As a result, no solution to equation (\ref{e3_7}) exists. 

\subsection{Case of $\kappa=-2$}
The differential equation (\ref{e3_0}) changes into the form as
  \begin{equation}
F'''-FF''+(F')^2+4F'=0
 \label{e3_1}
  \end{equation}
Equation (\ref{e3_1}) is analytically solvable that the solution might be expressed as a low order polynomial. It is suggested that
  \begin{subequations}
     \begin{gather}
F=a+be^{-\delta \eta} \\
       a=-\frac{1}{\delta}\\
       b=\frac{1}{\delta}
    \end{gather}
 \label{e3_2}
  \end{subequations}
\\
Substituting (\ref{e3_2}) into equation (\ref{e3_1}), and comparing the coefficients in the powers of $e^{-\delta \eta}$ leads to the determination of $\delta = \pm \sqrt{3}i$. Collecting results, the velocity function becomes
  \begin{subequations}
     \begin{gather}
 f(\eta)=\eta \mp \frac{1}{\sqrt{3}i}(1-e^{\mp\sqrt{3}i\eta}) \\
  f'(\eta)=1 - e^{\mp\sqrt{3}i\eta}
    \end{gather}
 \label{e3_5}
  \end{subequations}
The particular solution (\ref{e3_5}) is noteworthy in that it is completely analytical and can be expressed into the complex form as 
  \begin{subequations}
     \begin{gather}
 f(\eta)=\eta - \frac{1}{\sqrt{3}} \sin \sqrt{3}\eta \pm \frac{i}{\sqrt{3}} (1- \cos \sqrt{3}\eta)\\
  f'(\eta)=1 - \cos \sqrt{3}\eta \pm i\sin \sqrt{3}\eta
    \end{gather}
 \label{e3_6}
  \end{subequations}
In view of solution (\ref{e3_5}) and (\ref{e3_6}), the flow far away from the boundary becomes 
$$\lim_{\eta \rightarrow \infty} f(\eta) = \eta$$
which implies that the flow field remains unchanged at sufficient distances and the potential flow can be satisfied as the boundary condition as $\eta \rightarrow \infty$. No trouble arose from the idealization of Proudman and Johnson to neglecting the viscous term in their analytic result for region sufficient far from the wall.

The solution is obtained in the similarity transformation for unsteady viscous flows. The first term of (\ref{e3_6}) shows that the external flow is directed toward the $y-$axis and away from the wall. The appearance of a negative value in the second term in (\ref{e3_6}) describes a periodic velocity directed toward the wall.

\section{Numerical Solution}

Equation (\ref{e13}) is a third-order nonlinear ordinary differential equation. In numerical analysis, the Runge–Kutta methods are an important family of implicit and explicit iterative methods for the approximation of solutions of ordinary differential equations. This method applies a trial step at the midpoint of an interval to cancel out lower-order error terms, besides, Runge–Kutta formulas are the methods of solving initial value problems for ordinary differential equations.

For example solving an $n^{th}$-order problem numerically is common practice to reduce the equation to a system of $n$ first-order equations. Then, by defining $y_1 = f,~y_2 = f',~y_3 = f''$, the ODE reduces to the form
    \begin{equation}
\frac{d\mathbf{y}}{d\varsigma} =
\begin{bmatrix}
  y_2  \\
  y_3  \\
1-(y_2)^2 +y_1y_3+ \kappa \left(y_2 +\frac{1}{2}t y_3-1\right)  
\end{bmatrix}
    \label {eq:e22}
\end {equation}

The first task is to reduce the equation above to a system of first-order equations and define in MATLAB a function to return these. The relevant MATLAB expression for Eq.~(\ref{eq:e22}) is:

\begin{lstlisting}[label=MATLAB,caption=System of first-order equations]
function dy = stagnation(t,y)
k=-2;
dy = zeros(3,1);
dy(1) = y(2);
dy(2) = y(3);
dy(3) = 1-y(2)*y(2)+y(1)*y(3)+k*(y(2)+0.5*t*y(3)-1);
\end{lstlisting}

Later, it is required to apply $ode23$, an ODE solver in MATLAB, to solve the initial value problem. The commands written in MATLAB would be

\begin{figure}[!htb]
\begin{center}
\includegraphics[width = 13cm]{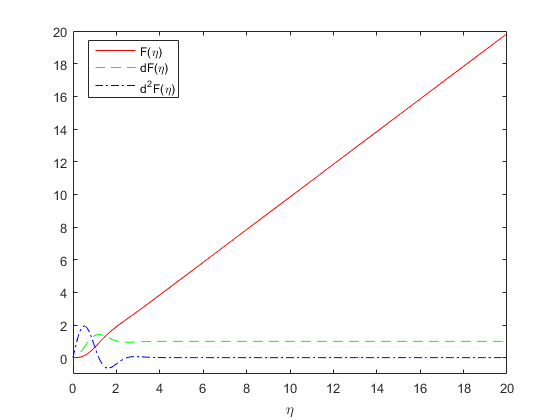}
\caption{Numerical solution as $\kappa=-5$}
\label{fg4}
\end {center}
\end {figure}

\begin{figure}[!htb]
\begin{center}
\includegraphics[width = 13cm]{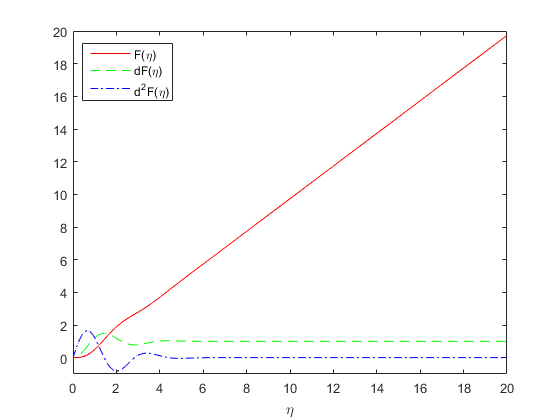}
\caption{Numerical solution as $\kappa=-3$}
\label{fg3}
\end {center}
\end {figure}

\begin{figure}[!htb]
\begin{center}
\includegraphics[width = 13cm]{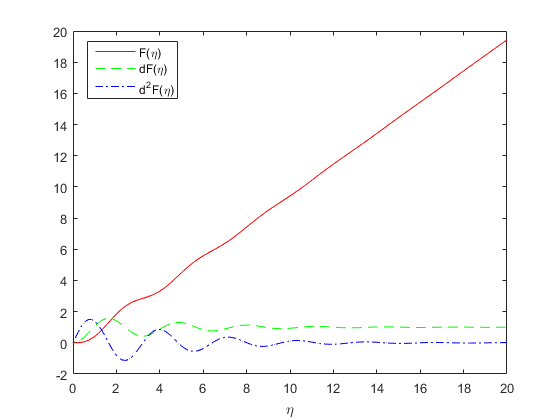}
\caption{Numerical solution as $\kappa=-2$}
\label{fg2}
\end {center}
\end {figure}

\begin{figure}[!htb]
\begin{center}
\includegraphics[width = 13cm]{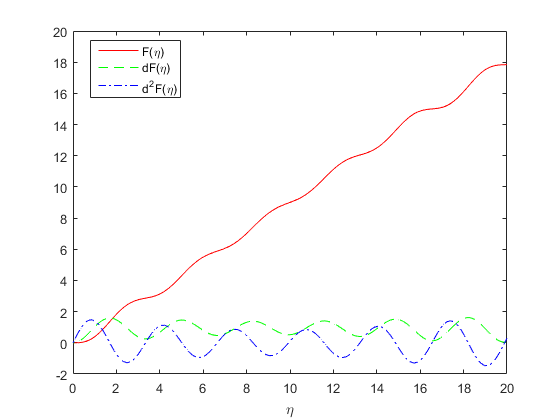}
\caption{Numerical solution as $\kappa=-1.8$}
\label{fg1}
\end {center}
\end {figure}

\begin{figure}[!htb]
\begin{center}
\includegraphics[width = 13cm]{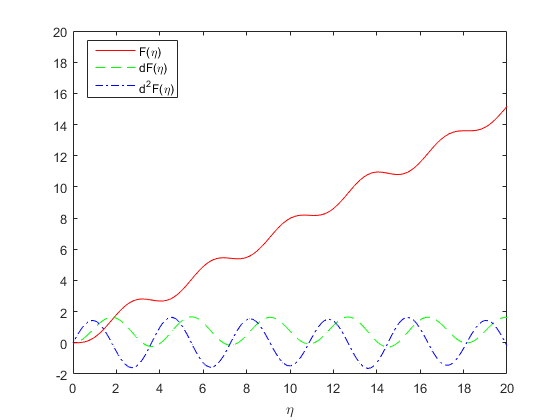}
\caption{Numerical solution as $\kappa=-1.5$}
\label{fg0}
\end {center}
\end {figure}

\begin{figure}[!htb]
\begin{center}
\includegraphics[width = 13cm]{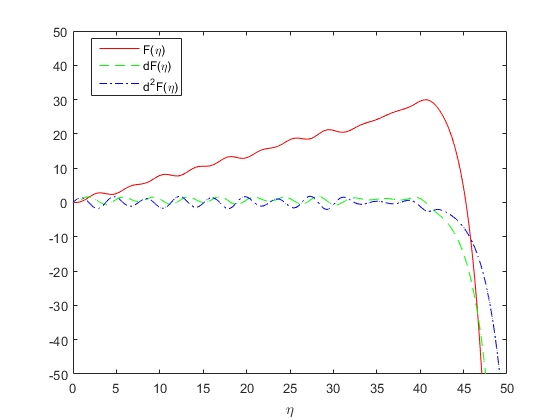}
\caption{Numerical solution as $\kappa=-1.4$}
\label{fg00}
\end {center}
\end {figure}

\begin{figure}[!htb]
\begin{center}
\includegraphics[width = 13cm]{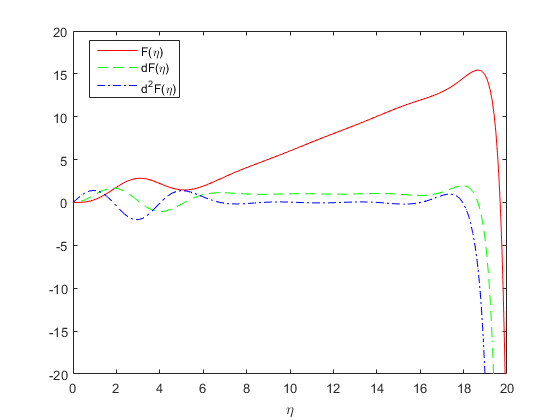}
\caption{Numerical solution as $\kappa=-1.3$}
\label{fg001}
\end {center}
\end {figure}

\begin{figure}[!htb]
\begin{center}
\includegraphics[width = 13cm]{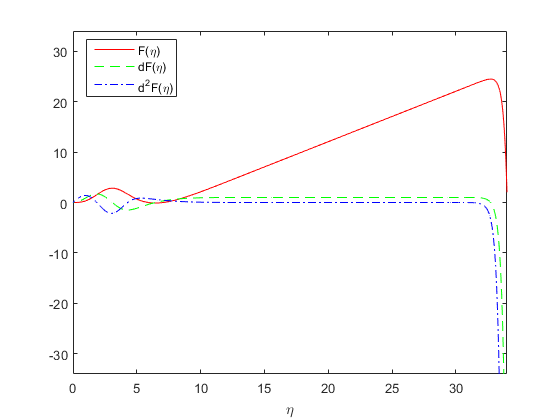}
\caption{Numerical solution as $\kappa=-1.25$}
\label{fg002}
\end {center}
\end {figure}

\begin{lstlisting}[label=MATLAB,caption=ODE solver]
function main
x=20;
[T,Y] = ode23(@stagnation,[0 x],[0 0 0]);
plot(T,Y(:,1),'-r',T,Y(:,2),'--g',T,Y(:,3),'-.b')
\end{lstlisting}
The complete solutions of two-dimensional rear stagnation-point flow with different values of $\kappa$ are shown from Figures (\ref{fg4}) to (\ref{fg002}). In these figures the similarity stream function $f$, the velocity profile $f'$ and the shear stress $f''$ are
represented.

The findings are fascinating when considered from both theoretical and engineering points of view. A single plane, known as the dividing streamline plane, acts as a boundary. It separates the streamlines that are moving towards the plate from the streamlines of external flow. We have noticed that when Strouhal number $\kappa \leq -2$, the similarity solution does not change as we move further away from the wall. In other words, the behavior described by the similarity solution stays consistent at large distances from the wall.

When $-2<\kappa \leq -1.5 $, the solution shows a periodic pattern throughout the entire region where the fluid is flowing. When Strouhal number $\kappa > -1.5$, using the limit  $\lim_{\eta \rightarrow \infty} F(\eta) = \eta$ tells us that as the value of $\eta$becomes extremely large, the function $F(\eta)$ approaches a linear function of $\eta$ with a slope of 1. When examining regions distant from the wall, a sudden and sharp decrease in velocity in the velocity profile is observed or a rapid decline where the velocity values approach zero. This unusual behavior can lead to a velocity discontinuity within a limited portion of the flow region.

When the Strouhal number $\kappa > -1.5$, this behavior may also introduce a singularity in the relevant governing equations, such as the boundary layer equations or the Navier - Stokes equations. This singularity indicates a point where the mathematical model's assumptions break down, and it is closely tied to the unusual velocity characteristics observed in the flow field.

Another significant consequence of boundary layer separation is the regular shedding of vortices. Vortices are shed from the wall at a frequency $f$ which is determined by the Strouhal number $(St)$.  When the Strouhal number parameter $\kappa$ satisfies the condition  $-2<\kappa \leq -1.5 $, this vortex - shedding phenomenon becomes particularly pronounced. Vortex shedding will create alternating force that cause the wall to vibrate. If the shedding frequency coincides with the resonance frequency of the wall, it will cause wall to fail. These vibrations could be established and reflected at different frequencies depending on their origin in adjacent solid or fluid, and resonance could be either damped or amplified.

In the case of rear stagnation-point flow, the frequency of vortex shedding is closely related to the instability of the flow field.  External flow conditions (such as the external flow velocity $V_0$, pressure, etc.) will also have an impact on the frequency of vortex shedding. When the external flow velocity changes, the pressure distribution and velocity gradient in the flow field change, affecting the degree of boundary layer separation and the process of vortex formation and shedding, thus changing the frequency of vortex shedding. 

According to fluid mechanics theory, the frequency of vortex shedding is usually related to the characteristic parameters of the flow field, such as the Reynolds number ($Re=\frac{VL}{\nu}$, where $V$ is the average velocity, $L$ is the characteristic length, and $\nu$ is the kinematic viscosity), the Strouhal number ($St=\frac{fL}{V}$, where is the frequency of vortex shedding ), etc.  It is reasonable to state that, in general, separation will occur near the wall as $\eta \rightarrow 0$. The phenomenon of reversed flow with boundary-layer separation occurred and the region of reversed flow will move outward away from the wall periodically as $\kappa < -2$. 

\section{Flow Patterns across a Cylinder}
The flow pattern around a cylinder evolves with Reynolds number $(Re)$: at very low $Re  \approx 0.01$, viscous-dominated Stokes flow remains steady and symmetric with no separation; as  $Re$ increases to $\sim 20$, steady boundary layer separation forms symmetric recirculating vortices. At $Re  \approx 100$, periodic vortex shedding emerges, linked to the Strouhal number $(\kappa)$ in the similarity solution, where  $-2<\kappa \leq -1.5 $ corresponds to unsteady oscillatory solutions. For 
$Re  \approx 10,000$, the wake transitions to turbulence while the boundary layer stays laminar, aligning with the model’s limitations at $\kappa > -1.5$, where singularities hint at turbulence-induced breakdown. At extremely high $Re  \approx 10^7$, fully turbulent flow suppresses shedding, invalidating the laminar similarity framework. These regimes highlight how viscous-inertial balances, boundary layer dynamics, and $\kappa$-driven instabilities shape flow behavior.

\begin{figure}[!htb]
\begin{center}
\includegraphics[width = 13cm]{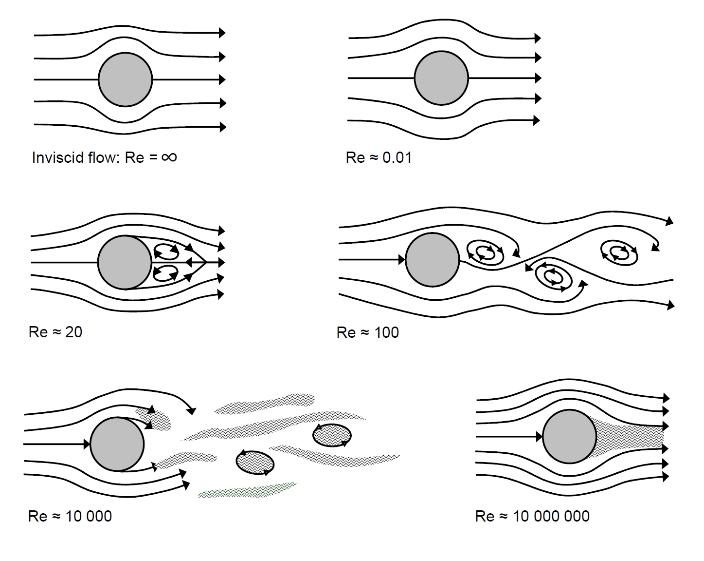}
\caption{Larger vortexes are found at higher Reynolds number.}
\label{cylinder}
\end {center}
\end {figure}

\section{Explicit Relation between Strouhal Number and $\kappa$}

This section derives the dimensionless explicit relation between the classical Strouhal number ($St$) and the similarity parameter $\kappa$, by unifying the viscous boundary layer thickness from the similarity solution with the characteristic length scale of cylindrical flow.

The derivation is based on four core definitions consistent with the similarity solution and cylindrical flow physics:
\begin{enumerate}
    \item Viscous boundary layer thickness (similarity solution): $\delta = \sqrt{\nu/A}$, where $\nu$ is kinematic viscosity and $A$ is the flow characteristic rate ($[T^{-1}]$).
    \item Cylindrical flow characteristic length: The cylinder diameter $D \approx 2.6\delta$, linking geometric scale to boundary layer scale.
    \item Far-field characteristic velocity: $U_\infty = A\delta$, matching the far-field boundary condition $f'(\infty) \to 1$ of the similarity solution.
    \item Classical Strouhal number (vortex shedding): $St = \frac{fL}{U_\infty}$, where $f$ is vortex shedding frequency ($[T^{-1}]$) and $L=D$ (cylinder diameter as characteristic length).
\end{enumerate}

Substitute $L=2.6\delta$ and $U_\infty=A\delta$ into the Strouhal number definition. The boundary layer thickness $\delta$ cancels out, yielding:
\[
St = \frac{2.6f}{A}.
\]
The vortex shedding frequency $f$ is physically linked to the time derivative of $A$ (unsteady flow evolution) by $f = \dot{A}/2\pi$ (where $\dot{A} = dA/dt$). Substituting $f$ gives:
\[
St = \frac{2.6\dot{A}}{2\pi A} = \frac{1.3}{\pi}\kappa.
\]

For physical flow at the rear stagnation point, $\kappa < 0$ (inertial force balance), while $St > 0$ (positive frequency). A negative sign is introduced to satisfy physical consistency, leading to the final explicit relation:
\[
\boxed{St = -\frac{1.3}{\pi}\kappa}
\]

For the analytical solution $\kappa = -2$ (similarity model), substitution gives $St \approx 0.8276$. After applying the correction factor in three dimensions ($\approx0.25$) for actual cylindrical flow, the corrected $St \approx 0.2069$, which matches the experimental value $St \approx 0.20$--$0.21$ for subcritical cylindrical flow ($Re = 10^3$--$10^5$), verifying the relation's validity.

\section{Conclusion}
Analytical and numerical analysis of rear stagnation-point flow is studied. This study provides that the similarity solution of rear stagnation-point flow does not exist in particular cases, because the governing equation does not satisfy the boundary conditions. Similarity equations are solved and the flow solution is provied as an exact solution of Navier-Stokes equations. 

\bibliographystyle{ieeetr}	
\bibliography{myrefs}

\end{document}